\pdfoutput=1

\documentclass[12pt,letterpaper]{article}
\usepackage{jheppub}
\usepackage{verbatim}

\usepackage{graphicx}
\usepackage{subfigure}
\input{epsf}
\usepackage{epsfig}
\usepackage{epstopdf}
\usepackage{amsthm}
\usepackage{mathrsfs}
\usepackage{verbatim}

\newcommand{\be}{\begin{equation}}
\newcommand{\ee}{\end{equation}}
\def\({\left(} \def\){\right)}
\def\[{\left[} \def\]{\right]}
\def\({\left(} \def\){\right)}

\def\tGamma{\tilde{\Gamma}}
\def\tG{\tilde{\Gamma}}

\title{Four-point function in the IOP matrix model}

\author[a]{Ben Michel,}
\author[a,b]{Joseph Polchinski,}
\author[b]{Vladimir Rosenhaus,}
\author[b,c]{and S.~Josephine Suh}

\affiliation[a] {Department of Physics, 
 University of California, Santa Barbara, CA 93106 }

\affiliation[b]{Kavli Institute for Theoretical Physics, 
 University of California, Santa Barbara, CA 93106}

\affiliation[c]{Department of Physics and Astronomy,
University of British Columbia, Vancouver, B.C.
}

\emailAdd{michel@physics.ucsb.edu}
\emailAdd{joep@kitp.ucsb.edu}
\emailAdd{vladr@kitp.ucsb.edu}
\emailAdd{sjsuh@phas.ubc.ca}
 
\abstract{
The IOP model is a quantum mechanical system of a  large-$N$ matrix oscillator and a fundamental oscillator, coupled through a quartic interaction. It was introduced previously as a toy model of the gauge dual of an AdS black hole, and captures a key property that at infinite $N$ the two-point function decays to zero on long time scales. Motivated by recent work on quantum chaos, 
we sum all planar Feynman diagrams contributing to the four-point function. We find that the IOP model does not satisfy the more refined criteria of exponential growth of the out-of-time-order four-point function.

}

\begin{document}
\maketitle

\section{Introduction}

Matrix models are useful toy  models of gauge theories and holography. Strongly coupled quantum field theories are difficult to understand directly, having a prohibitively large set of Feynman diagrams that must be summed. A good model should have a sufficiently small and well-organized set of diagrams,   allowing for the computation of the full planar correlation functions.  The diagrammatic structure should, however, be sufficiently nontrivial so as to capture the essential features of the bulk. 

The IP model \cite{IP} is a simple large-$N$ system of a harmonic oscillator in the $U(N)$ adjoint representation plus a harmonic oscillator in the $U(N)$ fundamental representation, coupled through a trilinear interaction. It has the same graphical structure as the 't Hooft model of two-dimensional QCD  \cite{tHooft:1974hx}. The IOP model \cite{IOP} is a more tractable variant of the IP model. It possesses the same degrees of freedom, but the trilinear interaction is replaced by one that is quartic in the oscillators but quadratic in the $U(N)$ charges. Building on ideas of \cite{Liu}, the IP and IOP models were introduced in \cite{IP, IOP} as toy models of the gauge theory dual of an AdS black hole. These models capture a key property of black holes: the long time decay of the two-point function at infinite $N$, but not at finite $N$ \cite{Maldacena:2001kr}. 

In this paper we compute the thermal four-point function in the IOP model in the planar limit. The motivation for studying the four-point function comes from recent work in quantum chaos and holography   \cite{KitaevNov, Kitaev, Maldacena:2015waa, Shenker:2013pqa, Shenker:2013yza,  Leichenauer:2014nxa, Roberts:2014isa, Roberts:2014ifa, Shenker:2014cwa,  Polchinski:2015cea, Berenstein:2015yxu, Hosura:2015ylk, Gur-Ari:2015rcq, Stanford:2015owe, Berkowitz:2016znt, Polchinski:2016xgd, Fitzpatrick:2016thx}. A signature of quantum chaos in a large-$N$ theory is the exponential growth in time of the connected out-of-time-order four-point function \cite{LO}. The growth rate is identified as a Lyapunov exponent. A black hole has a Lyapunov exponent of $2\pi T$ \cite{Kitaev, Shenker:2013pqa}, which is the maximal possible Lyapunov exponent \cite{Maldacena:2015waa}. The significance of the out-of-time-order four-point function as a diagnostic for the viability of a model of holography was recognized in \cite{KitaevNov}. 

In Sec.~\ref{sec:therm} we review the role of the two-point function as a diagnostic of thermalization. In Sec.~\ref{sec:chaos} we review the role of the out-of-time-order four-point function as a diagnostic of chaos. In Sec.~\ref{sec:SYK} we briefly mention the Sachdev-Ye-Kitaev model \cite{SY, Kitaev}, which was recently recognized to be maximally chaotic \cite{Kitaev}. We point out that the random coupling can, to leading order in $1/N$, be replaced by a quantum variable. 

In Sec.~\ref{sec:IP2pt} we review the calculation of the planar two-point function in the IP model. In Sec.~\ref{sec:IP4pt} we compute the planar four-point function. This involves summing ladder diagrams, which can only be done analytically in the limit of small adjoint mass, to which we restrict ourselves. 

In  Sec.~\ref{IOP2pt} we review the planar two-point function in the IOP model. In  Sec.~\ref{sec:IOP4pt} we compute the planar four-point function. Diagrammatically, the IOP model is more involved than the IP model. However, it has the advantage of allowing analytic computations for any adjoint mass. For both the IP and IOP models, we work in the limit that the mass of the fundamental is heavy, as compared to the temperature.

In the regimes considered, we find that the IP and IOP models are not chaotic. Some speculations on why this is so, and possible modifications of the models, are mentioned in Sec.~\ref{sec:Dis}.

\section{Thermalization, chaos, and large N}
\subsection{Thermalization}  \label{sec:therm}

Holography has provided useful insights into both strongly coupled field theories, as well as their gravity duals.  A  well-studied property of a black hole is its approach to equilibrium after a perturbation. A two-point function computed in a black hole background exhibits late time decay of the form \cite{Horowitz:1999jd, Danielsson:1999fa},
\be \label{2ptdecay}
\langle \phi(t) \phi(0)\rangle \sim e^{- c t/\beta}~,
\ee
where $c$ is an order-one constant and $\beta$ is the inverse temperature. The late time decay of the two-point function has a clear interpretation in the bulk: matter falls into the black hole, but classically nothing escapes. Computing subleading corrections in $G_N$ to (\ref{2ptdecay}) does not prevent the late time decay. 

As recognized in \cite{Maldacena:2001kr}, the late time decay to zero of a two-point function is inconsistent with the properties of  a finite entropy quantum mechanical system. On the field theory side, one thus has  the statement  that, to all orders in $1/N$, the two-point function decays to zero at late times, even though this property does not hold nonperturbatively in $1/N$. The two-point function $\langle \phi(t) \phi(0)\rangle_{\beta}$ can be regarded as the overlap between the states $\phi(0)|\beta\rangle$ and $\phi(t) |\beta\rangle$; its  decay is a probe of thermalization. Therefore, the large $N$ limit acts like a thermodynamic limit \cite{Liu}.

\begin{figure} 
\centering
\includegraphics[width=2.3in]{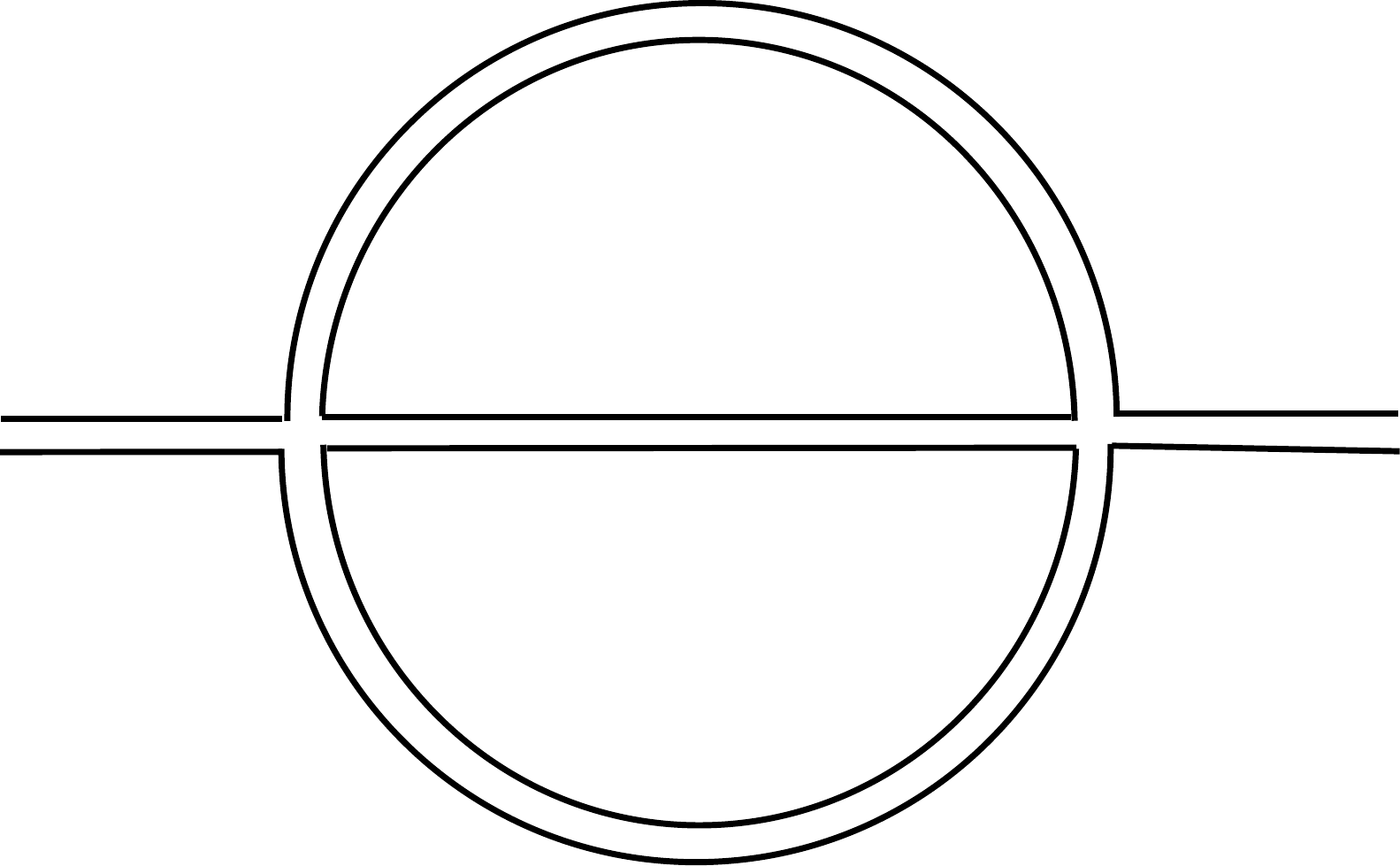}
\caption{The basic graphical unit of the Hamiltonian (\ref{eq:Liu}) studied in \cite{Liu}.} \label{fig:FLd}
\end{figure}

This late time breakdown of perturbation theory was studied in the context of matrix quantum mechanics in \cite{Liu}. Reducing Yang-Mills on a sphere in terms of spherical harmonics, one obtains a Hamiltonian whose essential features can be captured by considering just two interacting matrices. For instance,  \cite{Liu} considers two large $N$ matrices $M_1, M_2$ with a Hamiltonian,
\be \label{eq:Liu}
H = \sum_{i =1}^{2} \frac{1}{2} \text{Tr}(\dot{M_i}^2 + \omega_i^2 M_i^2) + \lambda \text{Tr}(M_1 M_2 M_1 M_2)~.
\ee
The  relevant diagrams for the decay of the two-point function are the sunset diagrams  shown in Fig.~\ref{fig:FLd}. 

\begin{figure} 
\centering
\includegraphics[width=2.4in]{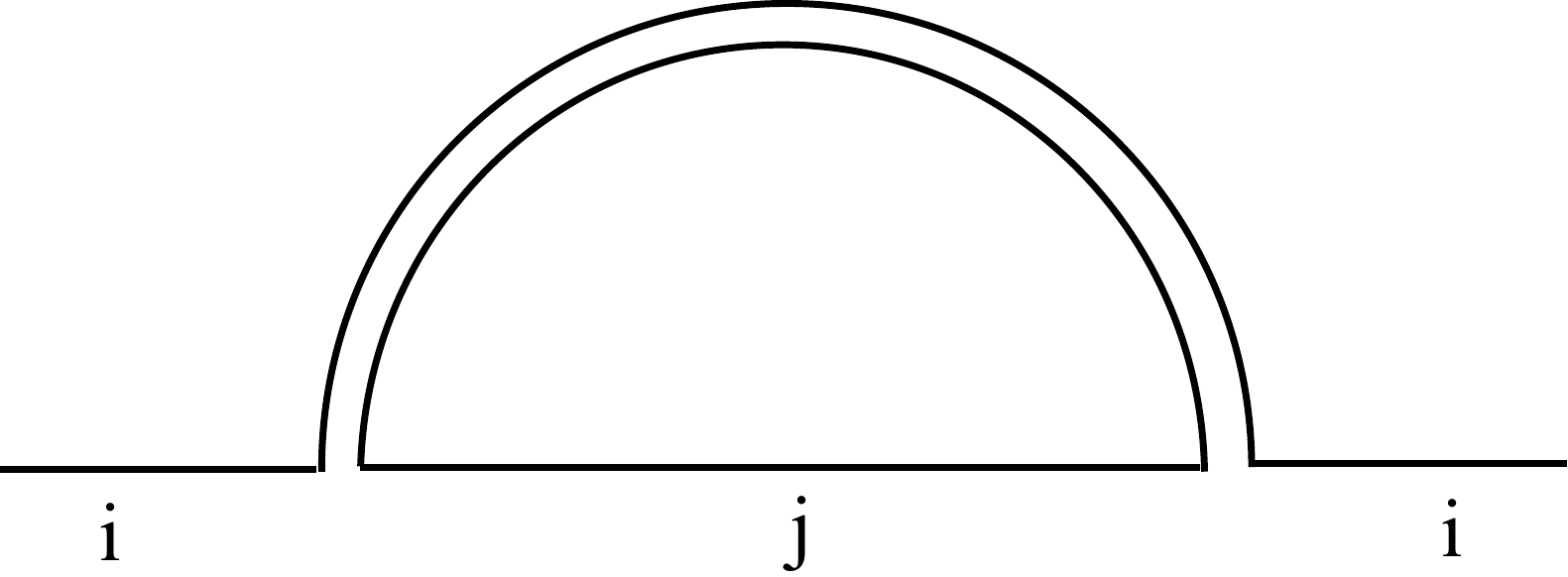}
\caption{The basic graphical unit of the IP model (\ref{HIP}) studied in \cite{IP}. It is like the diagram in Fig.~\ref{fig:FLd}, but cut in half. A single line is a fundamental, a double line is an adjoint. } \label{fig:IPd}
\end{figure}

The model (\ref{eq:Liu}) has the drawback of still being too complicated to allow the summation of all planar Feynman diagrams.  
The goal of \cite{IP} was to find a matrix model  that is more tractable, while still exhibiting the late time decay of the planar  two-point function. The IP model \cite{IP} is given by the Hamiltonian,
\be \label{HIP}
H_{IP}= m \text{Tr}(A^{\dagger} A) + M a^{\dagger}a +  g  a^{\dagger} X a~,
\ee
where $a_i$ is the annihilation operator for a harmonic oscillator in the fundamental of $U(N)$, while  $A_{ij}$ is the annihilation operator for an oscillator in the adjoint, and  $X_{i j}= (A_{i j} + A_{j i}^{\dagger})/\sqrt{2m}$.~\footnote{Since the highest term in the Hamiltonian (\ref{HIP}) is cubic, there is no ground state. This is cured by adding a stabilizing term, $ a^{\dagger} a(a^{\dagger} a -1)$, which vanishes in the relevant sectors $a^{\dagger} a =0,1$ \cite{IP}.} 
As we review in Sec.~\ref{sec:IP}, the planar two-point function can be found if one takes the mass of the fundamental to be large compared to the temperature, $M\gg T$. For a general mass $m$ for the adjoint, the planar Schwinger-Dyson equation for the two-point function can be solved numerically, exhibiting the desired late time exponential decay. In the limit of small mass for the adjoint, $m\rightarrow 0$, the two-point function can be found analytically, giving late time power law decay.

A variant of the IP model, the IOP model, was introduced in \cite{IOP},
\be \label{HIOP}
H_{IOP} = m \text{Tr}(A^{\dagger} A) + M a^{\dagger} a  + h a_i^{\dagger} a_l A_{i j}^{\dagger} A_{ j l}~.
\ee
This model has the feature that analytic computations are possible for any mass $m$. It again exhibits power law decay of the two-point function at long times.

\subsection{Chaos}  \label{sec:chaos}
Chaos in deterministic systems is understood as  aperiodic long-term behavior that exhibits sensitive dependence on initial conditions. Two points in phase space, characterized by a  separation $\delta x(0)$, will initially diverge at a rate,
\be \label{eq:div}
\delta x(t) = \delta x(0)\, e^{\kappa t},
\ee
 where $\kappa$ is the Lyapunov exponent.

For a number of reasons \cite{LesHouches}, there is no straightforward extension of chaos to quantum systems. 
In the semiclassical regime,  \cite{LO} gave an intuitive definition of chaos. Replacing the variation in (\ref{eq:div}) by a derivative, and noting that this is given by a Poisson bracket,
\be
\frac{\partial x(t)}{\partial x(0)} = \{x(t), p(0)\}~,
\ee
the generalization to quantum systems consists of replacing the Poisson bracket by a commutator.  The commutator is generally an operator, so seeing exponential growth requires taking an expectation value. The expectation value of the commutator in the thermal state will vanish, as a result of phase cancelations. A simple way to cure this is to consider the square of the commutator \cite{LO},~\footnote{The expectation value in (\ref{expxp}), and elsewhere, is in the thermal state. The Lyapunov exponent depends on the temperature: this is the familiar statement from classical chaos that  regions of phase space that do not mix have different Lyapunov exponents. If we were working in the microcanonical ensemble, then the energy would be conserved. Note also that the definition of the Lyapunov exponent that is being used is nonstandard, in that it is a local  Lyapunov exponent, rather than  involving a time average. } 
\be \label{expxp}
\langle \[x(t), p(0)\]^2 \rangle \sim \hbar^2 e^{2\kappa t}~.
\ee
Alternatively, one can consider the thermal expectation value of the commutator times the anticommutator; this will scale as $\hbar$. Either of these consist of sums of out-of-time-order four-point functions. The important point is that a chaotic system has an out-of-time-order four-point function that exhibits exponential growth.  The exponential growth persists until a time of order $- \kappa^{-1} \log \hbar$, at which point the commutator saturates at an order one value. 

For a large $N$ field theory, $1/N$ plays the role  of $\hbar$, and the classical limit is the infinite $N$ limit. For matrix models, such as the IP and IOP models, leading order in $1/N$ corresponds to keeping the planar Feynman diagrams. The criteria of chaos for evaluating the viability of a model is a powerful one, that was recognized in \cite{KitaevNov}. A good model of a strongly coupled gauge theory should having an exponentially growing out-of-time-order four-point function. Moreover, if it is to be dual to a black hole, the Lyapunov exponent  must match that of a black hole \cite{Kitaev, Shenker:2013pqa}.

\subsection*{Thermalization and chaos} \label{sec:TC}
There is generally an intimate connection between thermalization and chaos. In the context of classical systems, there is a precise version of this statement \cite{Arnold}, which we now review. 

Letting $A$ and $B$ be regions of phase space, occupying phase space volumes $\mu(A)$ and $\mu(B)$, respectively,  and letting $\phi_t$ denote time evolution, a dynamical system is said to be mixing if $\mu\[\phi_t A \cap B\] \rightarrow \mu(A)\cdot \mu(B)$ as $t\rightarrow \infty$, for all sets $A$ and $B$. In the notation of quantum mechanics, this is the statement that a system is mixing if the (connected) two-point function of any two operators decays to zero at late time. A system is defined to be ergodic if for every function $f$, the time mean of $f(x)$ is equal to the space mean of $f(x)$.  It is shown in \cite{Arnold} that mixing implies ergodicity, but ergodicity does not necessarily imply mixing. 

It is important to note that for a system to be mixing, the two-point function of all operators must decay. In fact, the IP and IOP models do not satisfy this criteria, as it is only the two-point functions of the fundamentals that exhibit late time decay.~\footnote{In other words, the IP and IOP models not fully thermalizing. If they had been, the absence of chaos in these models would have been puzzling.} The adjoints have a two-point function of a free harmonic oscillator; they have no self-interaction, and the interaction generated via the fundamentals is $1/N$ suppressed. Thus,  exponential growth of the out-of-time-order four-point function for the fundamentals is a more refined criteria than the decay of the two-point function of the fundamentals at long times.

\subsection{SYK model} \label{sec:SYK}
\begin{figure}
\centering
\includegraphics[width=2.3in]{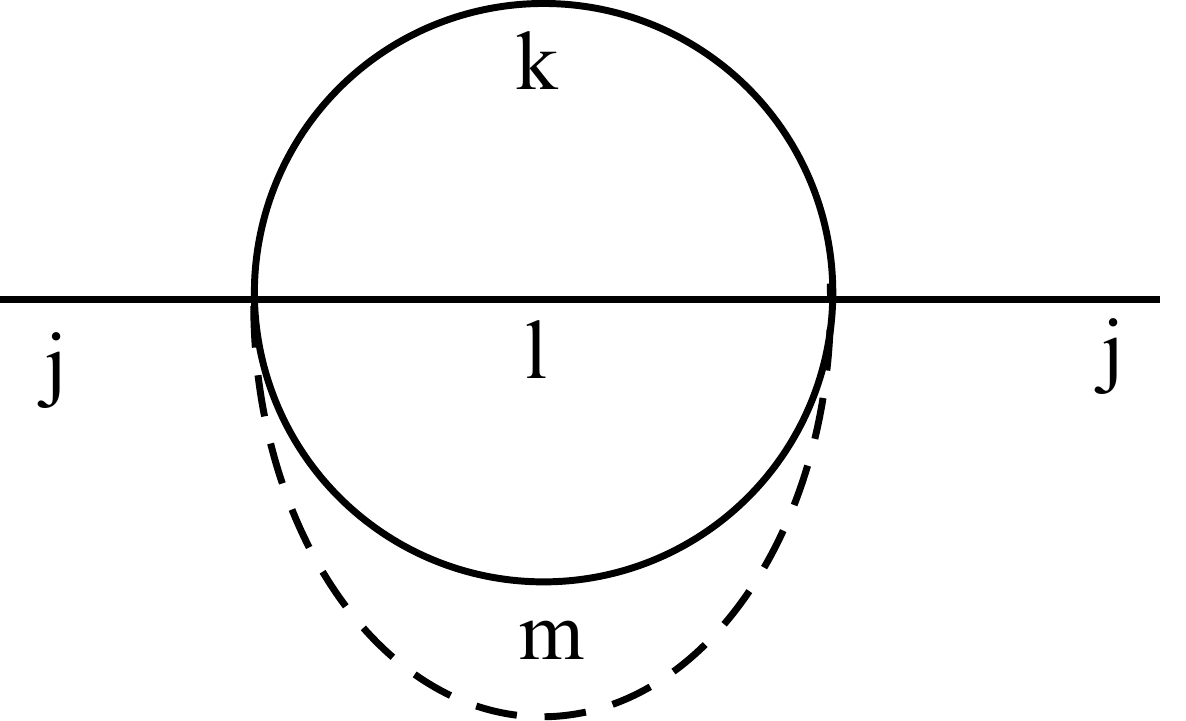}
\caption{The basic graphical unit of the SYK model (\ref{SYKH}). The solid lines are fermions $\chi_i$, the dotted line is  the coupling $J_{j k l m}$.}  \label{fig:SYKd}
\end{figure}

Kitaev has proposed a variant of the Sachdev-Ye model \cite{SY} as a model of holography \cite{Kitaev}. The SYK model consists of $N\gg 1$ Majorana fermions $\chi_i$ with a quartic interaction with random coupling $J_{j k l m}$, 
 \be \label{SYKH}
 H_{SYK} = \frac{1}{4!}\sum_{j,k,l,m=1}^N J_{j k l m}\ \chi_j \chi_k \chi_l \chi_m~, 
  \ee
 where couplings are drawn from the distribution, 
 \be \label{PJ}
 P(J_{j k l m}) \sim \exp(- N^3 J_{j k l m}^2/12 J^2)~,
 \ee
 giving a  disorder average of,
 \be \label{eq:JJJ}
 \overline{J_{j k l m}^2} = \frac{ 3! J^2}{N^3}\, , \  \ \ \ \ \overline{J_{j k l m}} = 0~.
 \ee
 Remarkably, one can analytically compute the disorder averaged large-$N$ correlation functions in the SYK model at finite temperature and strong coupling, $\beta J\gg 1$.  The  two-point function exhibits exponential late time decay, see \cite{SY, PG, Kitaev, Sachdev:2015efa}.    The out-of-time-order four-point function exhibits exponential growth \cite{Kitaev}, 
\be \label{4ptSYK}
   \langle \chi_i (t) \chi_j(0) \chi_i (t) \chi_j(0) \rangle \sim \frac{1}{N} e^{2 \pi t/\beta}~.
   \ee
For studies of the four-point function, see \cite{Kitaev, Polchinski:2016xgd, MStoappear, Ktoappear}.

An important aspect of the SYK model is the quenched disorder: if the coupling $J_{j k l m}$ where instead a fixed constant, there would be additional Feynman diagrams that would contribute at leading order in $1/N$. 
Here we simply point out that the disorder $J_{j k l m}$ can be replaced by a quantum variable, as the quantum corrections are $1/N^3$ suppressed.

\begin{figure} 
\centering
\includegraphics[width=6in]{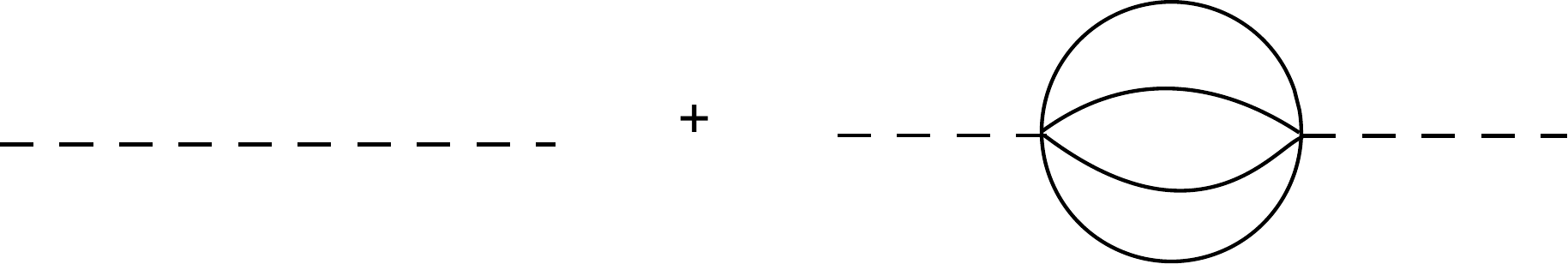}
\caption{The dashed lines indicate $J_{j k l m}$, while the sold lines are the fermions $\chi_i$. Treating $J_{j k l m}$ as a quantum field, the quantum corrections to the two-point function are suppressed by $1/N^3$.} \label{fig:dis}
\end{figure}

Recall that  the disorder averaged expectation value of an operator $O$ composed of the fields $\chi_i$ is, 
\be \label{eq:disAvg}
\overline{\langle O \rangle} = \int D J_{j k l m}\ e^{-  J_{j k l m}^2 N^3/12 J^2}\ \frac{\int D \chi_i\ O\ e^{- \int dt\, L}}{\int D \chi_i\ e^{- \int dt\, L}}~.
\ee
The interpretation of (\ref{eq:disAvg}) is that one first computes the expectation value $\langle O\rangle$ with some coupling $J_{ j k l m}$ drawn from the distribution (\ref{PJ}), and then averages over the $J_{j k l m}$.
If one were  to instead treat  $J_{j k l m}$ as a static  quantum variable, then the expectation value of $O$ would be given by,
\be \label{eq:eO}
\langle O \rangle = Z^{-1} \int D J_{j k l m}\ D \chi_i\ O \exp\(- N^3 J_{j k l m}^2/12 J^2 - \int dt\ L\)~.
\ee
In terms of Feynman diagrammatics, if $J_{j k l m}$ is a classical Gaussian-random parameter, then it has a two-point that is exactly $3! J^2/ N^3$. If instead $J_{j k l m}$ is a quantum variable, then its leading two-point function can be chosen to be $3! J^2/ N^3$, however this will receive quantum corrections, as shown in Fig.~\ref{fig:dis}. Thus, generally (\ref{eq:disAvg}) and (\ref{eq:eO}) are different. However, for the SYK model, the first quantum correction is suppressed by a factor of $1/N^3$: the loop diagram in Fig.~\ref{fig:dis} has two $J_{j k l m}$ propagators, giving a factor of $(3! J^2/ N^3)^2$ . So, at leading order in $1/N$, (\ref{eq:disAvg}) and (\ref{eq:eO}) are the same. Equivalently, the effective action for $J_{j k l m}$  is
\be
e^{- W[J_{j k l m}]} = \int D \chi_i\ \exp\( - J_{j k l m}^2 N^3/ 12 J^2 - \int dt\, L  \) = e^{ - J_{j k l m}^2 N^3/12 J^2} + \ldots~,
\ee
at leading order in $1/N$. 
Note that the structure of the vacuum is different depending on if $J_{j k l m}$ is quenched disorder or a quantum field: the vacuum loop scales like $N$, and receives a correction of the same order from interactions with $\chi_i$, as there is now a summation over the indices. This, however, is irrelevant for the purposes of connected correlation functions. 

The variable $J_{j k l m}$ is still not yet a standard quantum variable, due to the constraint that it be static. There are a few somewhat artificial ways to achieve this. One could add to the action a term $ \dot{J}_{j k l m} \phi$, where $\phi$ is some Lagrange multiplier field. A better option is to regard $J_{j k l m}$ as the momenta of harmonic oscillators for which the frequency is taken to zero.   Consider a harmonic oscillator with the standard Lagrangian, $(m \dot{x}^2 - m \omega^2 x^2)/2$. The Euclidean two-point function for the momentum is $\langle p(t) p(0)\rangle =m\omega e^{- \omega t}/2$. Now take the limit of $\omega \rightarrow 0$, so as to remove the time dependance. Letting $m \omega = 12 J^2 N^{-3}$, the momenta have the same two-point function as (\ref{eq:JJJ}).

\section{IP model} \label{sec:IP}
The IP model \cite{IP} is a quantum mechanical system, with a harmonic oscillator in the adjoint of $U(N)$  and a harmonic oscillator in the fundamental of $U(N)$, coupled through a trilinear interaction. The Hamiltonian for the IP model is given by (\ref{HIP}). The two-point function is found by summing rainbow diagrams (see Fig.~\ref{fig:IP2pt}) and is reviewed in Sec.~\ref{sec:IP2pt}. The four-point function is given by a sum of ladder diagrams (see Fig.~\ref{fig:ladder}), which we evaluate in Sec.~\ref{sec:IP4pt}. 

\subsection{Two-point function} \label{sec:IP2pt}
The bare zero temperature propagator for the fundamental is defined as,
\be
G_0(t) \delta_{ij} \equiv \langle T a_i(t) a_j^{\dagger}(0)\rangle e^{i M t}~.
\ee
Trivially, one has that,
\be \label{eq:G0}
G_0(t) = \theta(t), \ \ \ \ \ \ G_0(\omega) = \frac{i}{\omega +i \epsilon}~.
\ee

It will be assumed that fundamental has a large mass, $M\gg T$, where $T$ is the temperature. In this case, the bare finite temperature two-point function is the same as the bare zero temperature two-point function.

The adjoints have no self-interaction, and the backreaction from interactions with the fundamental is suppressed by $1/N$. Thus, the propagator for the adjoint is that of a free oscillator in a thermal bath,
\be
K(\omega) = \frac{i}{1-y}\left(\frac{1}{\omega^2-m^2 +i \epsilon} - \frac{y}{\omega^2-m^2-i\epsilon}\right)~,
\ee
where we have defined $y = e^{- m/T}$. It will be useful for later to note that in the limit that the adjoints become massless, $m\rightarrow 0$ ($y\rightarrow 1$), their two-point function becomes, 
\be \label{eq:K}
K(\omega) = \frac{2\pi }{1-y} \delta(\omega^2 - m^2)~.
\ee

\begin{figure} 
\centering
\includegraphics[width=4in]{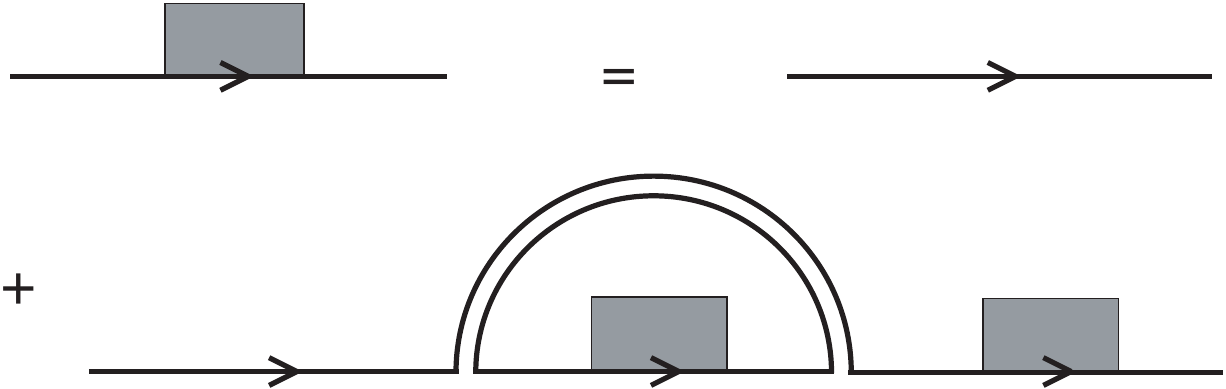}
\caption{The Schwinger-Dyson equation for the propagator $G(\omega)$ in the IP model, in the planar limit. Arrows point from creation operators toward annihilation operators.  A single line denotes the free propagator $G_0(\omega)$, a line with a shaded box is the dressed propagator $G(\omega)$, and a double line is the adjoint propagator  $K(\omega)$. Iterating generates a sequence of nested rainbow diagrams. } \label{fig:IP2pt}
\end{figure}

The planar two-point function for the fundamental is found by summing rainbow diagrams. The Schwinger-Dyson equation for the two-point function is given by (see Fig.~\ref{fig:IP2pt}):
\be \label{IPSD}
G(\omega) = G_0(\omega) + \lambda G_0(\omega) G(\omega)\int \frac{d\omega'}{2\pi} G(\omega') K(\omega - \omega')~,
\ee
where the 't Hooft coupling is $\lambda = g^2 N$.  In general, such an integral equation is difficult. However, the assumption that $M\gg T$ implies that $G(t) = 0$ for $t<0$. As a result, $G(\omega)$ has no poles in the upper half plane, allowing us to close the integration contour in (\ref{IPSD}) in the upper half plane $\omega'$ plane. Picking up the residues at $\omega' = \omega \pm m$, 
the Schwinger-Dyson equation turns into a difference equation,
\be
G(\omega) = \frac{i}{\omega + i \epsilon}\Big( 1- \frac{\lambda}{1-y} \frac{G(\omega)}{2 m} (G(\omega -m) + y G(\omega +m))\Big)~.
\ee
This can be solved numerically \cite{IP}, however to proceed analytically we take the limit of small adjoint mass and small 't Hooft coupling,
\be \label{eq:limit}
m\rightarrow 0, \ \ \ \ \nu^2 = \frac{2 \lambda}{m (1-y)} = \text{const.}
\ee
In this limit one finds \cite{IP},
\be \label{GIP}
G(\omega) = \frac{2 i}{\omega + \sqrt{\omega^2 - 2 \nu^2}}~.
\ee
Here the $\omega$ should really be an $\omega +i \epsilon$; we will generally suppress the $i \epsilon$, remembering that all the poles are in the lower half complex plane. 
The Fourier transform of the two-point function is a Bessel function,
\be\label{G2FT}
G(t) = \int \frac{d \omega}{2\pi} G(\omega) e^{- i \omega t} = \frac{\sqrt{2}}{ \nu t} J_{1}(\sqrt{2} \nu t)\, \theta(t)~.
\ee
We will later encounter integrals of a similar form, so we show (\ref{G2FT}) in some detail.  For positive times, the $\omega$ contour in (\ref{G2FT}) wraps around the branch cut stretching from $ - \sqrt{2} \nu< \omega <\sqrt{2} \nu$.~\footnote{Our $\omega$ integral was from $-\infty < \omega<\infty$. For positive time, we close the contour in the lower half plane. The branch cut is slightly below the real axis, and so is inside the contour. We can shrink the contour so that it hugs the branch cut. For negative times, the $\omega$ integral is closed in the upper half plane, and so gives zero. Also, our choice of location for the branch cut corresponds, for instance, to writing $\sqrt{\omega^2 - 2\nu^2} = \omega \exp\(\frac{1}{2} \log(1 - 2\nu^2/\omega^2)\)$ and taking the principal branch for the logarithm. } Using (\ref{GIP}) and moving the square root to the numerator, we rewrite (\ref{G2FT}) as, 
\be
G(t)= \frac{i}{\nu^2} \int \frac{d \omega}{2\pi} \(\omega - \sqrt{\omega^2 - 2 \nu^2}\)\, e^{- i \omega t}~.
\ee
The integral of the first term vanishes, while the second is twice a line integral, 
\be
G(t) =  \frac{1}{\nu^2} (1-e^{i \pi}) \int_{- \sqrt{2} \nu}^{\sqrt{2} \nu} \frac{d \omega}{2\pi}\,  \sqrt{2\nu^2 - \omega^2}\, e^{- i\omega t}~,
\ee
which gives (\ref{G2FT}). Now let us redo the calculation for the Fourier transform (\ref{G2FT}) slightly differently. Taking $G(\omega)$ in the form (\ref{GIP}) and changing variable to,
\be
x = \omega + \sqrt{\omega^2 - 2 \nu^2} , \ \ \ \ \ \omega = \frac{x^2 + 2\nu^2}{2 x}~,
\ee
gives, 
\be \label{G2FT2}
G(t) = \int \frac{dx}{2\pi}\frac{i}{x} \( 1- \frac{ 2\nu^2}{x^2} \) e^{ - \frac{i}{2} \(x + \frac{2 \nu^2}{x} \) t}~.
\ee
The $\omega$ contour in (\ref{G2FT}) that hugs the branch cut maps into an $x$ contour  that is a circle of radius $\sqrt{2} \nu$ and centered around the essential singularity at the origin. Using the integral representation of the Bessel function, 
\be \label{BessInt}
J_n (t) = \frac{i}{2\pi } \int dx\ x^{- n -1} e^{\frac{1}{2} t (x- x^{-1})}~,
\ee
where the contour circles clockwise around the origin, we have,
\be
G(t)  =\( J_0(\sqrt{2} \nu t) + J_2(\sqrt{2} \nu t)\) \theta(t)~,
\ee
which is equal to (\ref{G2FT}). At late time, $\nu t \gg 1$, the propagator decays as $G(t)\sim t^{-3/2}$.

\subsection{Four-point function}\label{sec:IP4pt}

\begin{figure}
\centering
\includegraphics[width=4.5in]{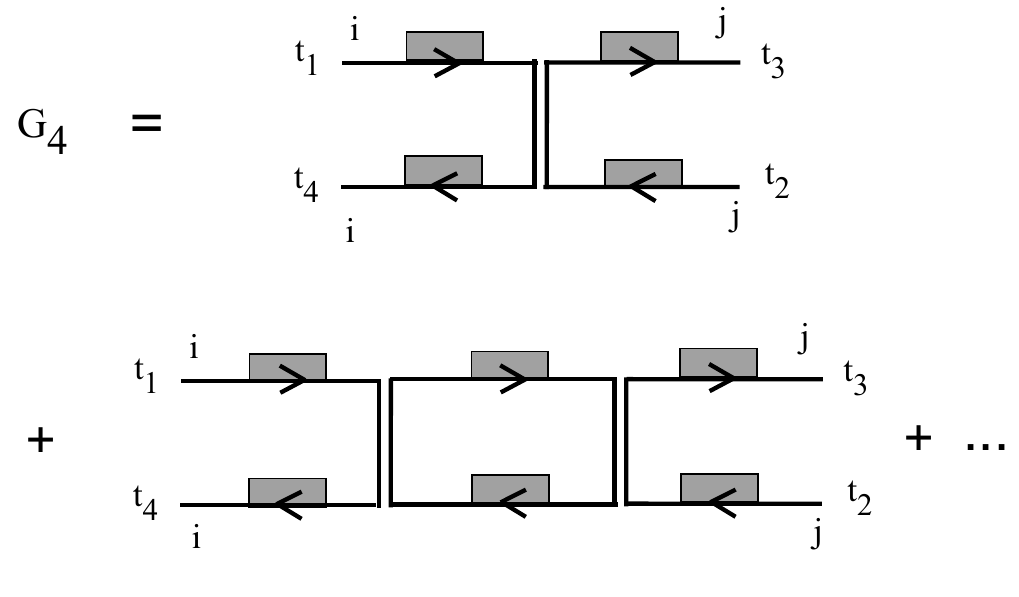}
\caption{The planar four-point function $G_4$ (\ref{eq:G4m1}) in the IP model. Ladders with $n=1$ and $n=2$ rungs are shown.  }  \label{fig:ladder}
\end{figure}

We now turn to the connected four-point function. In the planar limit, it consists of a sum of ladder diagrams, as shown in Fig.~\ref{fig:ladder}. The ingoing momenta are $\omega_1, \omega_2$, while the outgoing momenta are $\omega_3, \omega_4$.~\footnote{The ingoing momenta are drawn in Fig.~\ref{fig:ladder} as coming from the upper left and lower right in order for the diagram to look planar.} As in the case of the two-point function, to proceed analytically we must work in the limit specified in (\ref{eq:limit}). In this limit, the propagator for the adjoint is given by (\ref{eq:K}).

Consider the ladder diagram that consists of a single rung. It is given by,
\be \label{eq:single}
(-i g)^2 \int \frac{dp}{2\pi} G(\omega_1) G(\omega_1 - p) G(\omega_2) G(\omega_2+p) K(p)~.
\ee
Now inserting
\be
\delta(p^2-m^2) = \frac{1}{2m}\[\delta(p-m) + \delta(p+m)\]
\ee
into (\ref{eq:single}), evaluating the integral, and then taking the $m\rightarrow 0$ limit, yields for (\ref{eq:single}), 
\be
\frac{(-i g)^2}{m(1-y)}\ G(\omega_1)^2 G(\omega_2)^2~.
\ee

We  now sum all the ladder diagrams. As a result of the limit (\ref{eq:limit}), all the pieces appearing in the Feynman diagrams are on-shell. Defining $G_4(\omega_1, \omega_2, \omega_3, \omega_4) = \delta(\omega_1-\omega_3) \delta(\omega_2-\omega_4) G_4(\omega_1,\omega_2)$, and letting $n$ denote the number of rungs, we have
\be \label{eq:G4m1}
N G_4(\omega_1,\omega_2) = \sum_{n=1}^{\infty} \(\frac{-\lambda}{m(1-y)}\)^n \(G(\omega_1) G(\omega_2)\)^{n+1} = \frac{-\frac{\nu^2}{2}G(\omega_1)^2 G(\omega_2)^2}{1+ \frac{\nu^2}{2} G(\omega_1)G(\omega_2)} ~,
\ee
where $\nu$ was defined in (\ref{eq:limit}).

The Fourier transform of (\ref{eq:G4m1}) gives the position space four-point function,
\be \label{GFT}
N G_4(t_{31},t_{42}) = -\frac{\nu^2}{2}\int \frac{d\omega_1}{2\pi}\frac{d\omega_2}{2\pi}\ \frac{G(\omega_1) G(\omega_2)}{\frac{\nu^2}{2} + G(\omega_1)^{-1} G(\omega_2)^{-1}}\  e^{-i \omega_1 t_{31}}\ e^{-i \omega_2 t_{42}}~,
\ee
where we have defined $t_{31} \equiv t_3-t_1$, $t_{42} \equiv t_4-t_2$. In addition, $G(\omega)$ really denotes $G(\omega+ i\epsilon)$; we suppress the $i \epsilon$, remembering that,  if we are using $G$ in a time-ordered correlator,  all the poles are in the lower-half complex plane.

\subsubsection*{Free propagator}
The propagator entering the four-point function (\ref{GFT}) is given by (\ref{GIP}). As a warmup, it is useful to study (\ref{GFT}) with the free propagator (\ref{eq:G0}), rather than the dressed one. In this case we have, 
\be \label{G4a}
N \bar G_4(t_{31},t_{42}) = \nu^2 \int \frac{d\omega_1}{2\pi} \frac{d \omega_2}{2 \pi} \frac{1}{\omega_1 \omega_2} \frac{1}{\nu^2 - 2\omega_1 \omega_2}\ e^{- i \omega_1 t_{31}}\ e^{-i \omega_2 t_{42}}~.
\ee
Performing the $\omega_2$ integral, and closing the contour in the lower half plane, we pick up poles at $\omega_2 = 0$ and $\omega_2 = \nu^2/ 2\omega_1 $, 
\be \label{G4c}
N \bar{G}_4(t_{31},t_{42}) = -\theta(t_{42}) \theta(t_{31}) +  \theta(t_{42}) \int \frac{d \omega_1}{2\pi i}\frac{1}{\omega_1}\ e^{-i \(\omega_1 t_{31} + \frac{\nu^2}{2 \omega_1} t_{42}\)}~.
\ee
Using the integral representation of the Bessel function (\ref{BessInt}), we get,
\be \label{G4b}
N \bar G_4(t_{31},t_{42}) = \(J_0(\sqrt{2 t_{31} t_{42} }\nu) -1 \) \theta(t_{31}) \theta(t_{42})~.
\ee
Eq.~\ref{G4b} is the time-ordered four-point function, as evidenced by the theta functions. We can obtain the out-of-time-order four-point function by dropping the theta functions. 
In particular, setting $t_{31}= -t_{42} = t$ gives,
\be
N C(t) = I_0( \sqrt{2}\nu t) -1 ~.
\ee
In the limit $\nu t \gg 1$,
\be \label{ClargeT}
C(t) \rightarrow \frac{1}{2^{3/4} \sqrt{\pi \nu t} N} e^{\sqrt{2} \nu t}~.
\ee

By summing only a subset of the Feynman diagrams: the ladder diagrams with undressed propagators, we have found exponential growth in the out-of-time-order four-point function. While intriguing, using the free propagator is certainly not legitimate, as it violates unitarity; classically it would be equivalent to violating Liouville's theorem. However, before evaluating (\ref{GFT}) with the dressed propagator, it will be instructive to study (\ref{G4a}) a bit further. 

Returning to  (\ref{G4c}), and taking the limit of large $t_{31}, t_{42}$, we approximate the integral via the method of steepest descent (see Appendix~\ref{sec:steep}). This involves deforming the contour of integration in order for it to pass through the saddle point, at an angle so as to maintain constant phase. The saddle point of the exponent, 
\be
f(\omega_1) = \omega_1 t_{31} + \frac{\nu^2}{2 \omega_1} t_{42}~,
\ee
occurs at $\tilde \omega_1 = \pm \nu \sqrt{t_{42}/ 2 t_{31}}$. As we continue from a time-ordered four-point function, to an out-of-time-order four-point function, $t_{42}\rightarrow - t_{42}$, the saddle moves off of the real axis and onto the imaginary axis. At $t_{31} = -t_{42} =t$, the saddle is at $\tilde \omega_1 = \pm i \nu/\sqrt{2}$. The leading exponent in the integral in (\ref{G4c}) can therefore be approximated by,
\be
e^{- i t f(\tilde \omega_1)} = e^{\sqrt{2} \nu t}~,
\ee
reproducing (\ref{ClargeT}).

Let us also reproduce (\ref{G4b}) by returning to (\ref{eq:G4m1}) and computing the Fourier transform of each term before taking the sum. From (\ref{eq:G4m1}) and (\ref{eq:G0}) we have,
\be
\bar G_4(\omega_1, \omega_2) =- \sum_{n=1}^{\infty} \(\frac{\nu^2}{2}\)^n \frac{1}{(\omega_1 \omega_2)^{n+1}}~.
\ee
The Fourier transform gives,
\be \label{G4e}
\bar G_4(t_{31}, t_{42}) = \sum_{n=1}^{\infty}\(\frac{\nu^2}{2}\)^n \frac{(-t_{31} t_{42})^n}{(n !)^2}\ \theta(t_{31}) \theta(t_{42}) = \(J_0(\sqrt{2 t_{31} t_{42} }\nu) -1 \) \theta(t_{31}) \theta(t_{42})~,
\ee
where we have made use of the series definition of the Bessel function. 

The expression (\ref{G4e}) is easy to see directly in time-space. Since the free two-point function for the fundamental is simply $\theta(t)$ (\ref{eq:G0}), a ladder diagram with $n$ rungs will have $n+1$ propagators for the fundamentals on each of the two sides. For one such side we have a product of  theta functions, with the time insertions of the rungs integrated over. For the top side, 
\be
\int_{t_1}^{t_{3}} d t_{a_n} \ldots \int_{t_1}^{t_{a_3}} d t_{a_2 } \int_{t_1}^{t _{a_2 }} d t_{a_1} = \frac{1}{n!} t_{31}^n~,
\ee
and similarly a factor of $t_{42}^n/n!$ from the bottom side. Accounting for the coupling at each vertex, $-i g$, as well as the sum over indices, and the  factor of $m^{-1}(1-y)^{-1}$ coming from the adjoint propagator,  we recover the sum in (\ref{G4e}).

If we wish to form an out-of-time-order four-point function, for instance with $t_{42} <0$, then on the bottom edge of the ladder diagrams, time runs backwards: we must use a two-point function that is $\theta(-t)$ rather than $\theta(t)$. In addition, since time is running backwards on the bottom edge, the interactions come with a factor of $i g$, instead of $- i g$. This results in the elimination of the minus sign in the sum in (\ref{G4e}), and correspondingly gives exponential growth.

\subsubsection*{Dressed propagator}
We now return to the frequency-space four-point function (\ref{eq:G4m1}), and evaluate the Fourier transform (\ref{GFT}), this time using the full dressed propagator. Inserting the propagator $G(\omega)$ (\ref{GIP}) into (\ref{GFT}) gives, 
\begin{multline} \label{eq331}
N G_4(t_{31}, t_{42}) = - G(t_{31}) G(t_{42}) \\
+ 4 \int \frac{d \omega_1}{2\pi} \frac{d\omega_2}{2\pi} \frac{1}{2 \nu^2 - (\omega_1+ \sqrt{\omega_1^2-2\nu^2})(\omega_2 + \sqrt{\omega_2^2-2\nu^2})} e^{- i \omega_1 t_{31}}\ e^{- i\omega_2 t_{42}}~,
\end{multline}
where we have split off a $G(\omega_1) G(\omega_2)$ from  (\ref{eq:G4m1}), giving the first term in (\ref{eq331}). 
Changing integration variables to $x_i = \omega_i+\sqrt{\omega_i^2 - 2\nu^2}$ gives, 
\begin{multline} \label{GG4}
N G_4(t_{31}, t_{42}) = - G(t_{31}) G(t_{42}) \\
+ \int \frac{dx_1}{2\pi}\ \frac{dx_2}{2\pi}\ \(1-\frac{2\nu^2}{x_1^2}\)\(1-\frac{2\nu^2}{x_2^2}\) \ \frac{1}{2\nu^2 -x_1 x_2}\ e^{-\frac{i}{2}\(x_1 + \frac{2 \nu^2}{x_1}\)t_{31}}\ e^{-\frac{i}{2}\(x_2 + \frac{2 \nu^2}{x_2}\)t_{42}}~.
\end{multline}

Our goal is to see if (\ref{GG4}) exhibits exponential growth; if this does occur, it will be in the out-of-time-order regime, such as $t_{31} = - t_{42} = t$. We consider the late time limit,~\footnote{Since we are working in the planar limit, late time is still before the scrambling time, which scales as $\log N$.} and approximate (\ref{GG4}) via the saddle point method (Appendix~\ref{sec:steep}): we seek to deform the contours of integration of $x_1, x_2$ such that they pass through a saddle, at an angle such that the phase is constant. If we are away from the poles of the integrand, the saddle point occurs at $x_i = \pm \sqrt{2} \nu$, which clearly only gives oscillatory behavior. Now consider the regions at the poles of the integrand, at  $x_1 x_2 = 2 \nu^2$. This a peculiar region, as
\be \label{eq:333}
\omega = \frac{x}{2} + \frac{\nu^2}{x}
\ee
is invariant under $x\rightarrow 2\nu^2/x$.  Inserting this $x_2 = 2\nu^2/x_1$ into the exponent in (\ref{GG4}) , the exponent becomes,
\be
\exp\(- \frac{i}{2} \(x_1 + \frac{2 \nu^2}{x_1}\)(t_{31} + t_{42})\)~,
\ee
which does not give rise to the exponential growth indicative of chaos. Moreover, for $t_{31}=-t_{42}$, it simply vanishes.

\section{IOP model} \label{sec:IOP}
We now turn to the IOP model \cite{IOP}. Like the IP model, this is a quantum mechanical system, with a harmonic oscillator in the adjoint of $U(N)$  and a harmonic oscillator in the fundamental of $U(N)$. However, the interaction is now quartic in the oscillators  (\ref{HIP}), and quadratic in the $U(N)$ charges. The latter property makes the IOP model more analytically tractable than the IP model, although diagrammatically it is more involved.
As in the IP model, we consider the limit in which the fundamental is heavy, $M \gg T$. However, we can now obtain analytic results at any mass $m$ for the adjoint. 
We review the two-point function in Sec.~\ref{IOP2pt}, and  compute the four-point function in Sec.~\ref{sec:IOP4pt}.

\subsection{Two-point function} \label{IOP2pt}
\begin{figure}
\centering
\includegraphics[width=5in]{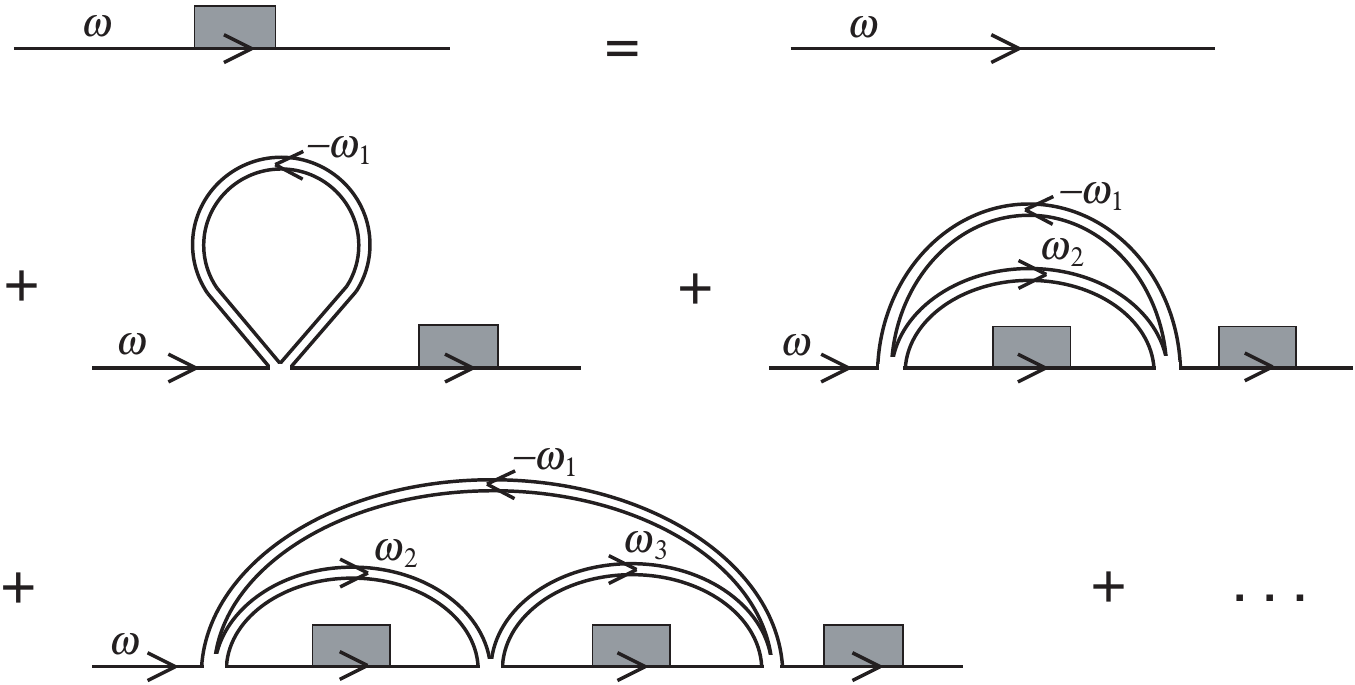}
\caption{Planar Feynman graphs for the fundamental propagator $G(\omega)$ (\ref{eq:Gfull2}) in the IOP model. The shaded rectangles mark the full planar propagators. Arrows point from creation operators toward annihilation operators. The graphs for $n$ = 0, 1, 2 are shown. }  \label{fig:IOP2pt}
\end{figure}

The bare propagator for the fundamental is the same as in the IP model (\ref{eq:G0}). The propagator for the adjoint is  that of free harmonic oscillator in a thermal bath, defined by $ L(t) \delta_{i l} \delta_{j k} = \langle T  A_{i j}(t) A_{k l}^{\dagger}(0)\rangle$, and giving,
\be \label{IOPad}
L (\omega)= \frac{i}{1-y}\[\frac{1}{\omega-m + i \epsilon} - \frac{y}{\omega-m - i \epsilon}\] ~.
\ee
The Schwinger-Dyson equation for the planar two-point function for the fundamental is (see Fig.~\ref{fig:IOP2pt}),
\begin{eqnarray}
G(\omega) &=& G_0(\omega) + G_0(\omega) G(\omega) \sum_{n=0}^{\infty} S_n(\omega)~, \\
S_n(\omega) &=& (-i h N)^{n+1} \int \frac{d^{n+1} \vec{\omega}}{(2\pi)^{n+1}} L(-\omega_1) \prod_{l=1}^{n} \[G(\omega - \omega_{l+1} - \omega_1) L(\omega_{l +1})\]~.
\end{eqnarray}
As $G$ only has poles in the lower-half plane, we can close the $\omega_i$ integrals in the lower-half plane and pick up residues only from $L$. This leads to an algebraic equation for $G$, with the solution \cite{IOP},
\be \label{eq:Gfull2}
G(\omega) =\frac{2 i}{\lambda + \omega + \sqrt{(\omega- \omega_+)(\omega - \omega_-)}}, \ \ \ \ \ \ \omega_{\pm} = \lambda \frac{1 +y \pm 2 \sqrt{y}}{1 -y}~,
\ee
where the 't Hooft coupling is $\lambda = h N$. The propagator has a branch cut from $\omega_- $ to $\omega_+$, leading to late-time power law decay, $t^{-3/2}$. 

\subsection{Four-point function} \label{sec:IOP4pt}
We now turn to the  four-point function in the planar limit.  The connected four-point function is found by summing ladder-like diagrams, shown in Fig.~\ref{fig:ladder4pt}, where each ``rung'' of the ladder is found by summing an infinite number of diagrams, like the ones shown in Fig.~\ref{fig:ladder2}. We warm up by computing the four-point function in the limit of small adjoint mass $m$, before doing the calculation for arbitrary $m$. 

\subsubsection*{Small adjoint mass}
\begin{figure}
\centering
\subfigure[]{
\includegraphics[width=6in]{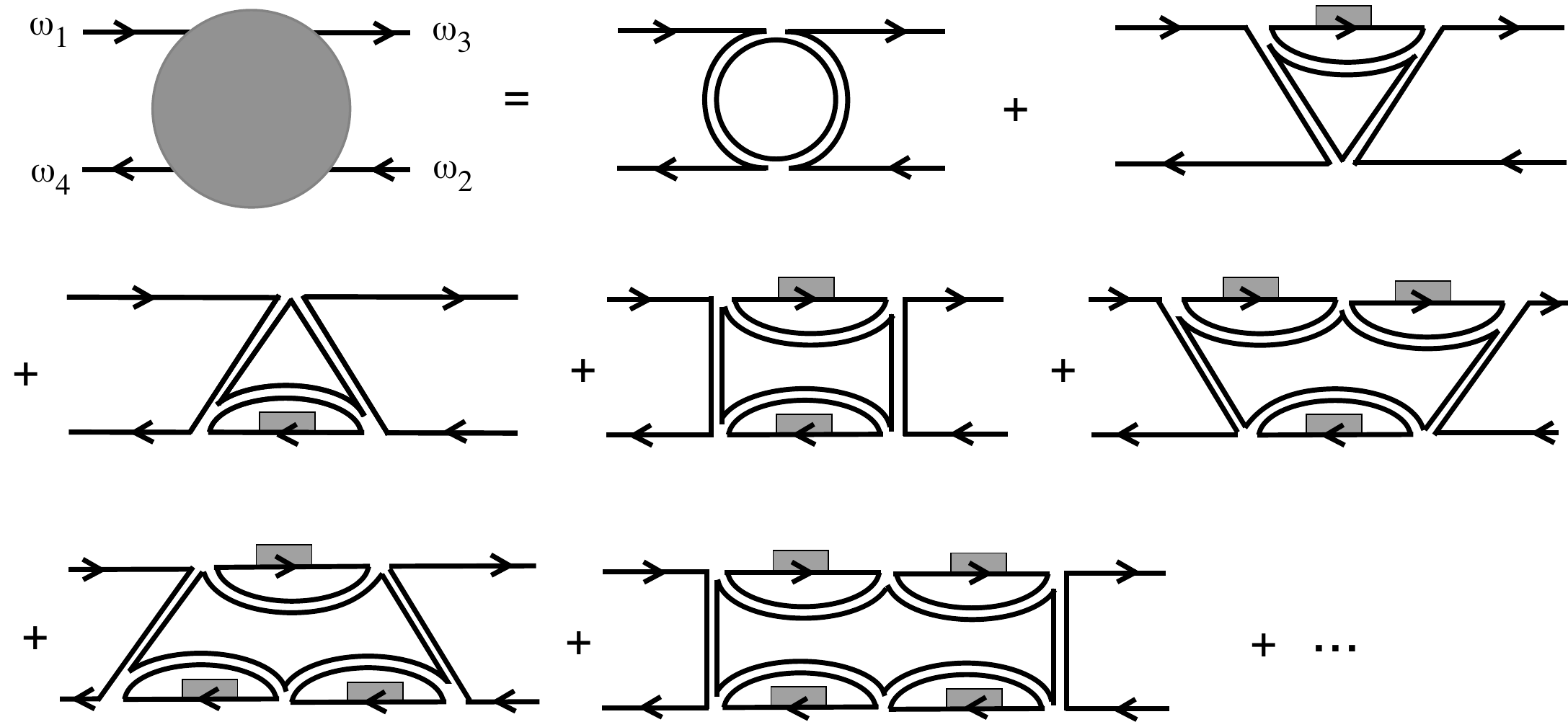}
}
\caption{Planar diagrams contributing to a ``rung'' $\Gamma$ in the IOP model. Diagrams with $n,m=0,1,2$ are shown. }  \label{fig:ladder2}
\end{figure}
We start with the limit $m\rightarrow 0$. In particular, 
\be
m\rightarrow 0, \ \ \ \ \ \kappa \equiv  \frac{\lambda}{1-y}~,
\ee
where $\kappa$ is held constant. In this limit, the two-point functions for the adjoint (\ref{IOPad}) and the fundamental (\ref{eq:Gfull2}) become,
\begin{eqnarray} 
L(\omega) &=& \frac{1}{1-y} 2\pi \delta(\omega-m)~,\\ \label{eq:Glimit}
G(\omega) &=& \frac{2 i}{\omega + \sqrt{\omega(\omega - 4 \kappa)}}~.
\end{eqnarray}
To compute the four-point function, we first sum the diagrams  shown in Fig.~\ref{fig:ladder2}, to get
\be
\Gamma(\omega_1,\omega_2, \omega_3, \omega_4) = \frac{(2\pi)^2}{N} \Gamma(\omega_1, \omega_2)  \delta(\omega_{13}) \delta(\omega_{24})~,
\ee
where $\omega_{i j} \equiv \omega_i - \omega_j$ and,
\be \label{eq:Gamma}
\Gamma(\omega_1, \omega_2) =\sum_{n,m=0}^{\infty} G(\omega_1)^n G(\omega_2)^m (- i \kappa)^{n+m+2} = \frac{-\kappa^2}{(1+i\kappa G(\omega_1))(1+i\kappa G(\omega_2))}~,
\ee
where the index $n/m$ labels the number of intermediate fundamental propagators on the top/bottom edge. As in the IP model, as a result of the $m\rightarrow 0$, all intermediate  propagators are on-shell. The  four-point function is given by the ladder-like sum of the $\Gamma$ (see Fig.~\ref{fig:ladder4pt}),
\be \label{eq:G4}
N G_4 (\omega_1, \omega_2)\! =\! \sum_{k=1}^{\infty}\! \Gamma(\omega_1, \omega_2)^k (G(\omega_1) G(\omega_2))^{k+1} =\! \frac{ G(\omega_1) G(\omega_2)}{1- \Gamma(\omega_1,\omega_2) G(\omega_1) G(\omega_2)}\, - G(\omega_1) G(\omega_2)~.
\ee
Inserting  (\ref{eq:Gamma}) into (\ref{eq:G4}) gives the frequency-space four-point function $G_4(\omega_1,\omega_2, \omega_3, \omega_4) = (2\pi)^2 \delta(\omega_{13})\delta(\omega_{24}) G_4(\omega_1,\omega_2)$ where,
\be \label{eq:G42}
N G_4(\omega_1,\omega_2) =\frac{-\kappa^2 G(\omega_1)^2 G(\omega_2)^2}{1+ i \kappa(G(\omega_1) + G(\omega_2))}~.
\ee 

Like in the IP model, we find exponential growth in the out-of-time-order four-point function if we only sum the diagrams containing the free propagator: (\ref{eq:G42}) with (\ref{eq:G0}) and $t_{31} = - t_{42} = t$ gives a four-point function $  \sim N^{-1} \exp(2 \kappa t)$ for large $t$. 

Now consider (\ref{eq:G42}) with (\ref{eq:Glimit}). The position-space four-point function is thus,
\be \label{mjk}
N G_4 (t_{31}, t_{42})\! =\! \int \frac{d \omega_1}{2\pi} \frac{ d\omega_2}{2\pi} \frac{- \kappa^2 G(\omega_1) G(\omega_2)}{G(\omega_1)^{-1} G(\omega_2)^{-1} + i \kappa (G(\omega_1)^{-1} + G(\omega_2)^{-1})}\ e^{- i \omega_1 t_{31}} e^{- i\omega_2 t_{42}}~.
\ee
Changing integration variables to $x_i = \omega_i + \sqrt{\omega(\omega - 4\kappa)}$, (\ref{mjk}) becomes
\be \label{eq:413}
N G_4 (t_{31}, t_{42})\!=\!\int\! \frac{ d x_1}{2\pi} \frac{d x_2}{2\pi} \frac{(x_1 - 4\kappa)}{(x_1 - 2\kappa)^2}\!\frac{(x_2 - 4 \kappa)}{(x_2 - 2 \kappa)^2} \frac{-4 \kappa^2}{x_1 x_2 - 2 \kappa(x_1 + x_2)}\ e^{- i\frac{x_1^2}{2(x_1 - 2\kappa)} t_{31}} e^{- i\frac{x_2^2}{2(x_2 - 2\kappa)} t_{42}}~.
\ee
We approximate the integral by taking the limit of large time separations, and looking for saddle points which could give rise to exponential growth. Picking up the pole at $x_1 x_2 = 2\kappa(x_1 + x_2)$, the exponent becomes, 
\be \label{eq:414}
\exp\(- i \frac{x_1^2}{2(x_1 - 2 \kappa)} (t_{31} + t_{42})\)~.
\ee
Like in the IP model, there is no exponential growth.

\subsubsection*{Arbitrary adjoint mass}
\begin{figure} 
\centering
\includegraphics[width=4.5in]{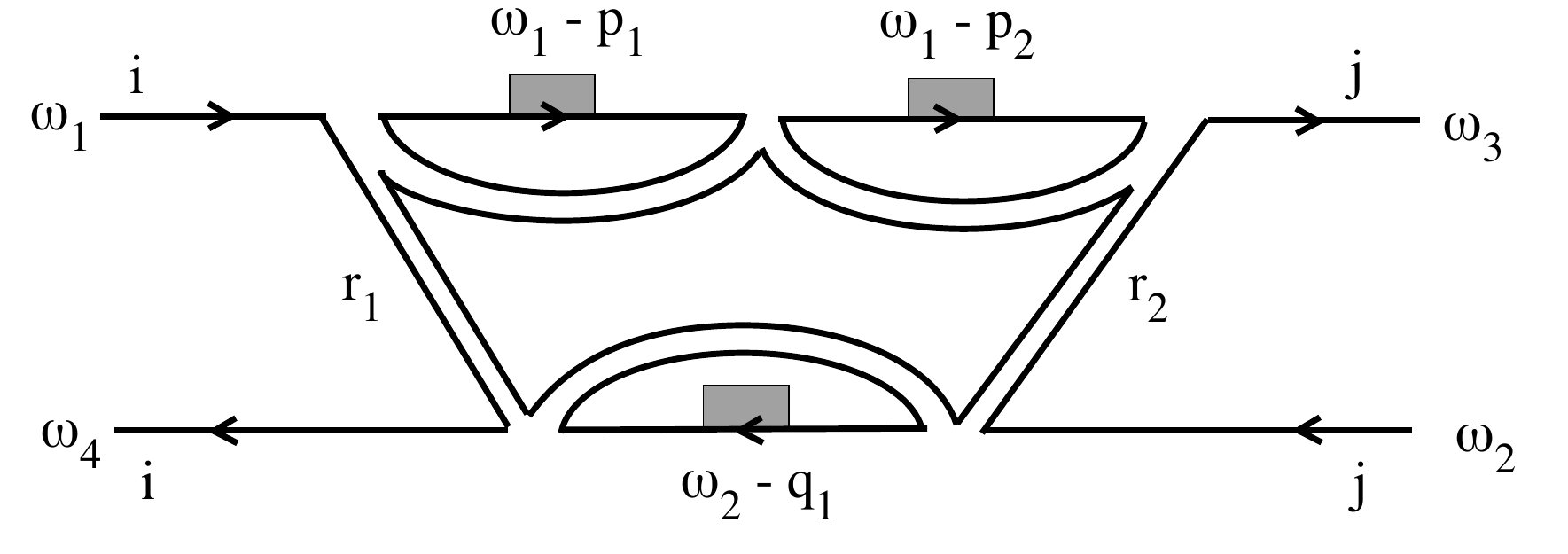}
\caption{One of the diagrams entering $\Gamma$ in Fig.~\ref{fig:ladder2}, given by $n=2, m=1$ in (\ref{uio}). } \label{fig:ladder3}
\end{figure}

We now compute the four-point function, with the adjoints taking arbitrary mass $m$. The Feynman diagrams contributing to ``rung'' $\Gamma$ are shown in Fig.~\ref{fig:ladder2}. A term in this sum, having  $n$ fundamental propagators on the upper edge and $m$ on the lower, is given by,
\be\label{uio}
(-i\lambda)^{n+m+2} \int \frac{d^n \vec{p}}{(2\pi)^n}\frac{d^m \vec{q}}{(2\pi)^m}  \frac{d r_1}{2\pi} 
L(r_1) L(r_2)\prod_{i=1}^n G(\omega_1-p_i)  L(r_1+p_i) \prod_{j=1}^m G(\omega_2-q_j) L(r_2+q_j)~,
\ee
where the ingoing frequencies are $\omega_1,\omega_2$, the outgoing frequencies are $\omega_3, \omega_4$, and we have defined $r_2=r_1+\omega_1-\omega_3$, and suppressed an overall factor of $N^{-1}$. In Fig.~\ref{fig:ladder3} the $n=2, m=1$ diagram from Fig.~\ref{fig:ladder2} is shown in more detail.
Since $G(\omega_1-p_i)$ has poles in the upper half $p_i$ plane, we close the contour in the lower half plane. Similarly for the $q_i$ integral. This gives for (\ref{uio}),
\be \label{ytr}
\frac{(-i\lambda)^{n+m+2}}{(1-y)^{n+m}} \int \frac{dr_1}{2\pi}\ L(r_1) L(r_2)\ G(\omega_1+r_1-m)^n\,G(\omega_2+r_2-m)^m~.
\ee
Evaluating the integral over $r_1$ by closing the contour in the upper half-plane, (\ref{ytr}) becomes,~\footnote{The adjoint propagator $L$ is given by (\ref{IOPad}). We denote the epsilon for $L(r_1)$ by $\epsilon_1$, and for $L(r_2)$ by $\epsilon_2$. Without loss of generality, we choose $\epsilon_2> \epsilon_1$. One can equally well choose $\epsilon_2<\epsilon_1$; this can be seen by rewriting  $(\omega_1-\omega_3-i (\epsilon_2-\epsilon_1))^{-1} = (\omega_1-\omega_3+i (\epsilon_2-\epsilon_1))^{-1} + 2\pi i \delta(\omega_1-\omega_3)$, and noting that $\delta(\omega_1-\omega_3) ( G(\omega_3)^n G(\omega_2)^m - G(\omega_1)^n G(\omega_4)^m) = 0$. }
\begin{multline} \label{xcv}
 \frac{i y (-i\lambda)^{n+m+2} }{(1-y)^{n+m+2}} \[ G(\omega_1)^n G(\omega_4)^m \(\frac{1}{\omega_1 - \omega_3 +  i \epsilon_1 + i \epsilon_2} - \frac{y}{\omega_1 - \omega_3 + i \epsilon_1- i\epsilon_2}\)\right. \\
 \left.  \ +\ G(\omega_3)^n G(\omega_2)^m \( \frac{1}{\omega_3 -\omega_1 +  i \epsilon_2+i\epsilon_1} - \frac{y}{\omega_3 - \omega_1 - i \epsilon_1+i\epsilon_2}\) \]~.
\end{multline}
To sum over all the  diagrams  contributing to $\Gamma$ (see Fig.~\ref{fig:ladder2}), we must sum (\ref{xcv}) over $n,m$ from $0$ to infinity. This gives $\Gamma = y \tGamma$ where, 
\be \label{eq:tGamma}
-i \tGamma(1,2,3,4) =  \frac{ z(1,4)}{\omega_1 - \omega_3 + i \epsilon} -\frac{y\,z(1,4) + (1-y)\,z(2,3)}{\omega_1 - \omega_3 - i \epsilon}, 
\ee
where we have defined,
\be \label{zjl}
z(j,l) = \frac{-\kappa^2}{(1+i \kappa G(\omega_j)) (1+ i \kappa G(\omega_l))}~,
\ee
 and have simplified notation to denote $\omega_j$ by  $j$, and recall that $\kappa \equiv \lambda/(1-y)$. One can also rewrite $\tGamma$ in (\ref{eq:tGamma}) as,
\be \label{GGGG}
y\tG =  y^2\, z(1,4)\, 2\pi \delta(\omega_1 - \omega_3) + y(1-y) \[ \frac{i\ z(1,4)}{\omega_1- \omega_3 + i \epsilon} - \frac{i\ z(2,3)}{\omega_1- \omega_3 - i\epsilon}\] ~,
\ee
which, recalling that $\omega_1+ \omega_2 = \omega_3 + \omega_4$, is manifestly symmetric under $\omega_1 \leftrightarrow \omega_2, \omega_3\leftrightarrow \omega_4$. 
\begin{figure}
\centering
\includegraphics[width=4.5In]{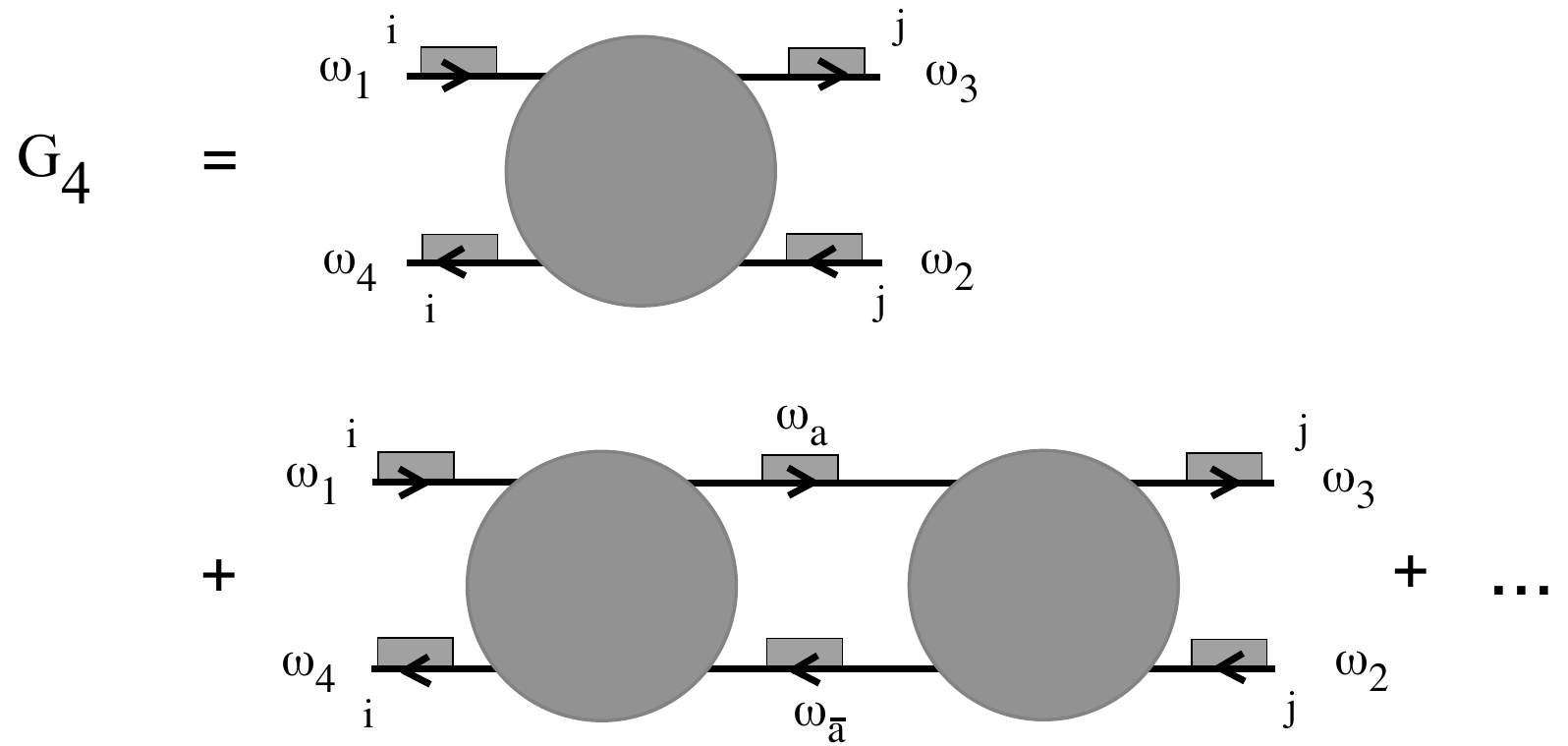}
\caption{The planar four-point function consists of ladders formed by gluing together the diagrams shown in Fig.~\ref{fig:ladder2}.}\label{fig:ladder4pt}
\end{figure} 

Attaching external propagators to (\ref{eq:tGamma}) gives the first term in the sum for the four-point function shown in Fig.~\ref{fig:ladder4pt}. The second term requires gluing two of the $\tGamma$ together, 
\be \label{shg}
(\tGamma \times \tGamma) (1,2,3,4) \equiv \int \frac{d \omega_a}{2\pi}\, G(a) G(\bar{a})\,\tGamma(1,\bar{a},a,4)\, \tGamma(a,2,3, \bar{a})~,
\ee
where $\omega_{\bar{a}} = \omega_a + \omega_4 - \omega_1$. Performing the integral  in (\ref{shg}) by closing the $\omega_a$ contour in the upper-half plane gives,
\be
\tGamma \times \tGamma =G(2) G(3)\, z(2,3)\, \tGamma + \frac{i (1-y) }{\omega_1 - \omega_3 + i \epsilon}\, G(1)G(4)\, z(1,4)\Big( z(1,4) -z(2,3)\Big)~,
\ee
where both the $\tGamma \times \tGamma $ on the left, and the $\tGamma$ on the right, are functions of the external $\omega_i$.

Let us simplify notation and let  $(\tGamma)^2$ denote  $\tGamma \times \tGamma$, defined  by (\ref{shg}). We define $(\tGamma)^n$, arising from gluing $n$ of the $\tGamma$ together, in an analogous fashion,
\be
(\tG)^n(1,2,3,4) \equiv \int \frac{d\omega_a}{2\pi}\, G(a) G(\bar{a})\, (\tG)^{n-1}(1,\bar{a},a,4)\, \tG(a,2,3,\bar{a})~.
\ee
We  compute $(\tGamma)^n$ iteratively, by gluing together $(\tGamma)^{n-1}$ and $\tGamma$. The result is, 
\be \label{Gnth}
(\tG)^n = G(2) G(3)\, z(2,3) (\tG)^{n-1} + \frac{i (1-y)}{\omega_1 - \omega_3 + i \epsilon}  \Big(G(1) G(4) z(1,4) \Big)^{n-1}  \Big( z(1,4) - z(2,3)\Big)~,
\ee
where we have for convenience expressed $(\tG)^n$ in terms of $(\tG)^{n-1}$. Next, we sum all the $(\tG)^n$. Denoting the sum by $S$, 
\be 
S =  \sum_{n=1}^{\infty} (\Gamma)^n~, 
\ee
where recall that $\Gamma = y \tGamma$, and separating off the $n=1$ term and using (\ref{Gnth}) for the rest, we get,
\be
S \Big( 1- y G(3) G(2) z(2,3) \Big) = \Gamma +  \frac{i}{\omega_1 - \omega_3 + i \epsilon} \frac{y^2(1-y) G(1) G(4) z(1,4) \Big(z(1,4) -z(2,3)\Big)}{1 - y G(1) G(4) z(1,4)} ~.
\ee
The four-point function is given by $S$, with external propagators attached. 

Thus, the connected  four-point function for the IOP model in the planar limit is, 
\begin{multline}\label{eq:G4S}
N G_4 (1,2,3,4) =\\ A(1,2,3,4)\, 2\pi \delta(\omega_1 + \omega_2 - \omega_3 -\omega_4)
\( y z(1,4) 2\pi \delta(\omega_1-\omega_3) + y(1-y) \frac{i B(1,2,3,4)}{\omega_1 - \omega_3+i \epsilon} \) ~, 
\end{multline}
where
\begin{eqnarray}
A(1,2,3,4) &=& \frac{G(1) G(2) G(3) G(4)}{1 - y G(2) G(3) z(2,3)}~, \\
B(1,2,3,4) &=& \frac{z(1,4) - z(2,3)}{1 - y G(1) G(4) z(1,4)}~,
\end{eqnarray}
where $j$ denotes the frequency $\omega_j$, the propagator $G(i)$ for the fundamental is given by (\ref{eq:Gfull2}), the constant $y$ is the Boltzmann factor $y= e^{-m/T}$ where $m$ is the mass of the adjoint and $T$ is the temperature,  and $z(j,l)$ was defined in (\ref{zjl}) and is a function of  $G(j)$, $G(l)$, and $\kappa = \lambda/(1-y)$, where $\lambda$ is the 't Hooft coupling. In the limit of small adjoint mass $m$ ($y\rightarrow 1$), the first term in (\ref{eq:G4S}) survives and reproduces the earlier result (\ref{eq:G42}). The out-of-time-order four-point function does not exhibit exponential growth with time, for reasons similar to those seen in the small adjoint mass limit (\ref{eq:413}, \ref{eq:414}); see Appendix~\ref{appendixB}.

\section{Discussion} \label{sec:Dis}
The absence of exponential growth in the out-of-time-order four-point function implies that the IOP model is not chaotic. In fact, there is a  heuristic way to understand the absence of chaos in the IOP model. The interacting part of the Hamiltonian (\ref{HIOP}) can be written as,
\be \label{QQ}
H_{\text{int}}= -h\, q_{l i} Q_{i l}, \ \ \ \ \text{    }\ \ \ q_{l i} = -a_i^{\dagger} a_l,\ \   Q_{i l} = A^{\dagger}_{i k} A_{k l}~.
\ee
As a result of the absence of self-interactions for the adjoints, combined with the assumption of large fundamental mass $M\gg T$, the number of fundamentals is time-independent and,
\be
a_i(t) = e^{-i h Q_{i l} t} a_l (0)~.
\ee
Since $Q$ is a hermitian matrix, it has real eigenvalues, and so the  norm of the  $a_i$ operators does not grow. 

If we relax the assumption that $M\gg T$, the above argument is no longer applicable, though this may not be sufficient to make the model chaotic. Heuristically, chaos is associated with rapid growth. As we evolve a fundamental, it is emitting and absorbing adjoints. Since the adjoints have no self-interaction, and conversion of an adjoint into two fundamentals is suppressed by $1/N$, the only way for the adjoints to continue evolving in between  emissions and absorptions is if they interact with fundamentals in the thermal bath. 

It may be useful to modify the IOP model, so as to have several flavors of fundamentals. Also, the interaction (\ref{QQ}) can written in terms of the quadratic Casimirs, $- h q\cdot Q = \frac{1}{2} h \text{Tr}(q^2 + Q^2 - (q+Q)^2)$, allowing a computation of the two-point function at finite $N$ through a sum over Young tableaux \cite{IOP}. One could study the four-point function in this way as well.

\acknowledgments
We thank 
O.~Aharony, D.~Berenstein, T.~Grover, A.~Kitaev, Z.~Komargodski, A.~Lawrence, D.~Roberts, M.~Smolkin, S.~Shenker, B.~Shraiman, M.~Srednicki, D.~Stanford, A.~Zhiboedov for helpful discussions. The work of BM was supported by the NSF Graduate Research Fellowship Grant No. DGE-1144085.  The work of JP and VR was supported by  NSF Grants PHY11-25915 and PHY13-16748. The work of JS was supported by a KITP graduate fellowship under grant PHY11-25915,  by the Natural Sciences and Engineering Research Council of Canada, and by grant 376206 from the Simons Foundation.

\appendix

\section{Steepest descent} \label{sec:steep}
In this appendix, we review some aspects of evaluating integrals by the method of steepest descent, see e.g.\  \cite{Bender}. 
Consider an integral of the form,
\be \label{IntSteep}
\int dz\ g(z)\, e^{-i t f(z)}~, \ \ \ \ \ \ t\gg1~,
\ee
where the integral is evaluated along some contour. For now, let $g(z), f(z)$ be smooth functions; we will discuss later how to relax this assumption. 
Since $t\gg1$, the integrand generically undergoes rapid oscillations which cancel out. The idea will be to deform the contour of integration so as to follow a path for which the phase remains constant. As long as we do not cross any singularities, we are free  deform the contour. Splitting $f(z)$ into a real and imaginary part, 
 \be
 f(z) = u(z) + i v(z)~,
 \ee
we need to deform the contour to follow a path of constant $u(z)$. The most relevant region of the integrand is one in which the real part is maximized. Letting $z=a+ i b$,
\be
\frac{\partial v}{\partial a} = \frac{\partial v}{\partial b} = 0~.
\ee
As a result of the Cauchy-Riemann equations, this amounts to finding the saddle points, $f'(z)=0$. Therefore, the prescription for approximating (\ref{IntSteep}) is  to focus on the vicinity of the dominant saddle point, and choose a direction for the contour that moves away from the saddle point so as to maintain constant phase $u(z)$. 

As an  example, consider the integral representation of the Bessel function,
\be \label{K0}
K_0(t) =\frac{1}{2} \int_{-\infty}^{\infty} dx \frac{e^{- i t x}}{\sqrt{1+x^2}}
\ee
This has a branch cut, $x \in (-i\infty, -i) \cup (i,i\infty) $.  We perform a change of variables, $x = \sinh u$, thereby bringing (\ref{K0}) into the form (\ref{IntSteep}),
\be \label{K02}
\frac{1}{2} \int_{-\infty}^{\infty} du \exp(- i t \sinh u)~.
\ee
Extermizing $f(u) = \sinh u$, the saddle points are at $u = \pm \pi i/2$. The line of constant phase passing through the saddle points is one that runs along the imaginary axis. We deform the contour so that it runs along $-\infty<u <- i \pi/2$. Moving downward from $u_0=-i \pi/2$ is a direction of steepest descent. In the vicinity of the saddle, 
\be
f(u) = f(u_0) + \frac{f''(u_0)}{2} (u-u_0)^2 + \ldots~.
\ee
Defining a new variable $z$ as $u =u_0 -i z$, (\ref{K02}) becomes,
\be \label{steep2}
\int_0^{\infty} d z \exp\( -t -t \frac{z^2}{2}\) = \sqrt{\frac{\pi}{2t}} e^{-t}~,
\ee
which is the correct large $t$ expansion of $K_0(t)$. 

We have so far discussed approximating (\ref{IntSteep}) by the behavior near the saddle point. There are several contexts in which other regions may be relevant. If the contour has endpoints, then one must analyze the behavior near the endpoints. Additionally, if $g(z)$ has singularities, then one must analyze the integrand near those regions as well. In particular, it may happen that there is no way to deform the contour into the relevant steepest descent contour, without passing through singularities. If the singularity of $g(z)$ is a simple pole, then we may simply deform through it, picking up the contribution of the pole. If, instead, $g(z)$ has a branch cut or an essential singularity, we must analyze the integrand in the vicinity of these regions. 

For instance, consider again approximating (\ref{K0}), but without changing variables. In this case, $g(x) =(1+x^2)^{-1/2}$ and $f(x) = x$. The exponential has no saddle points, so we focus on the regions where $g(x)$ is large: near $x=\pm i$. We integrate along a direction running parallel to the imaginary axis, as we still need to maintain constant phase for the exponent. Letting $x = - i - \rho i$, with $\rho \ll 1$ so that $\sqrt{1+x^2} \approx \sqrt{2\rho}$, (\ref{K0}) is approximated by,
\be \label{qwe}
\frac{e^{-t}}{\sqrt{2}} \int_0^{\infty} d\rho \frac{e^{- \rho t}}{\sqrt{\rho}}~,
\ee
where we have extended the range of integration to infinite $\rho$, as its contribution is negligible. Evaluating (\ref{qwe}) gives (\ref{steep2}). 

\section{Four-point integral} \label{appendixB}
The four-point function for the IOP model is,
\be \label{G4ps}
G_4(t_1,t_2, t_3, t_4) = \int \frac{d\omega_1}{2\pi} \frac{d\omega_2}{2\pi} \frac{d\omega_3}{2\pi}\, G_4(\omega_1,\omega_2, \omega_3,\omega_4)\, e^{- i \omega_1 t_{41} - i\omega_2 t_{42} - i \omega_3 t_{34}}~,
\ee
where $\omega_4 = \omega_1 + \omega_2 - \omega_3$ and $G_4(\omega_1,\omega_2, \omega_3,\omega_4)$ is given by (\ref{eq:G4S}). 

Our eventual interest is the out-of-time-order four-point with time separations $t_{41} = 0$, $t_{34} =-t_{42}= t$ and large $t$.  At large $t$, the exponent in (\ref{G4ps}) undergoes rapid oscillations as $\omega_2, \omega_3$ are varied. Since the exponent clearly has no saddle point, the only regions which could lead the four-point function to grow exponentially are those in which $G_4(\omega_1, \omega_2, \omega_3, \omega_4)$ is singular. We thus hold $\omega_1$ fixed, and scan over $\omega_2, \omega_3$, looking for regions in which the frequency-space four-point function is divergent. The relation between $\omega_2$ and $\omega_3$ where this occurs then determines the form of the exponent in (\ref{G4ps}), which can then be written just as a function of $\omega_2$. This function may  have saddles, which will either lead to an oscillatory exponent or a growing one. 

There are two terms in $G_4(\omega_1, \omega_2, \omega_3, \omega_4)$  given by (\ref{eq:G4S}). Consider the first of these, 
\be \label{szx}
y \frac{ z(1,4) G(1) G(2) G(3) G(4)}{1 - y G(2) G(3) z(2,3)} 2\pi \delta(\omega_1 -\omega_3)~,
\ee
where, as before, $G(j)$ denotes $G(\omega_j)$. It is convenient to rewrite (\ref{szx}) as 
\be \label{B3}
y G(2) G(3) \frac{1}{z(2,3)^{-1} G(2)^{-1} G(3)^{-1} - y} 2\pi \delta(\omega_1 - \omega_3)~,
\ee
where from (\ref{zjl}) we have that,
\be \label{B4}
z(j, l)^{-1} G(j)^{-1} G(l)^{-1} = -\frac{1}{\kappa^2} (G(j)^{-1} + i \kappa) (G(l)^{-1} + i \kappa)~.
\ee
It is convenient to rewrite the propagator (\ref{eq:Gfull2}) as,
\be \label{B5}
G(j) = \frac{2 i}{x_j}, \ \ \ \ \ x_j = \kappa(1-y) + \omega_j + \sqrt{\omega_j^2 - 2(1+y) \kappa\,\omega_j + \kappa^2 (1-y)^2}~.
\ee
Inverting the relation between $\omega_j$ and $x_j$, 
\be \label{oj}
\omega_j = \frac{x_j}{2} \( 1+ \frac{ 2\kappa y}{x_j - 2 \kappa}\)~.
\ee
Notice that (\ref{oj}) has a symmetry; $\omega_j$ is invariant under,
\be \label{eq:Inv}
x_j - 2\kappa \rightarrow \frac{4 \kappa^2 y}{x_j - 2\kappa}~.
\ee
This is analogous to the invariance seen in the IP model, see (\ref{eq:333}), as well as in the IOP model earlier, for $y=1$.
Now, the term (\ref{B3}) is singular when the denominator vanishes. Substituting (\ref{B4}, \ref{B5}), this occurs at $- x_2 x_3 + 2 \kappa (x_2 + x_3) - 4\kappa^2 (1-y) = 0$, which is,
\be \label{B8}
x_3 = 2\kappa\(1 + \frac{ 2\kappa y}{x_2 - 2\kappa}\)~.
\ee
As a result of the invariance (\ref{eq:Inv}), this implies $\omega_2 = \omega_3$. This is the same as what was seen for the IOP model at $y=1$, see (\ref{eq:414}).  Thus, the exponent in (\ref{G4ps}), as a function of $\omega_2$, is  oscillatory, and the same holds at the location of its saddle. 

Now consider the second term in $G_4(\omega_1, \omega_2, \omega_3, \omega_4)$, which is, 
\be \label{red}
 y(1-y) \frac{ G(1) G(2) G(3) G(4)}{1 - y G(2) G(3) z(2,3)}\frac{z(1,4) - z(2,3)}{1- yG(1) G(4) z(1,4)} \frac{i}{\omega_1 - \omega_3 + i \epsilon}~.
\ee
It is convenient to rewrite (\ref{red}) as,
\be \label{tyu}
y(1-y) \frac{z(2,3)^{-1} - z(1,4)^{-1}}{(z(2,3)^{-1} G(2)^{-1} G(3)^{-1} -y)(z(1,4)^{-1} G(1)^{-1} G(4)^{-1} -y)} \frac{i}{\omega_1 - \omega_3 + i \epsilon}~.
\ee
We  regard (\ref{tyu}) as a function of $\omega_2, \omega_3$, where recall that $\omega_4 = \omega_1 + \omega_2 - \omega_3$. The nontrivial singularities in (\ref{tyu}) arise from $(z(2,3)^{-1} G(2)^{-1} G(3)^{-1} -y)=0$, which as shown in (\ref{B8}) implies $\omega_2 = \omega_3$, or from $(z(1,4)^{-1} G(1)^{-1} G(4)^{-1} -y) =0$, which again gives $\omega_2 = \omega_3$. Thus, there is no regime of exponential growth for the four-point function.


\end{document}